\documentclass[aps,twocolumn,superscriptaddress]{revtex4}
\usepackage{epsfig}
\usepackage{times}
\usepackage{color}
\begin{document}

\title{Evolution of public cooperation in a monitored society with implicated punishment and within-group enforcement}

\author{Xiaojie Chen}
\email{xiaojiechen@uestc.edu.cn} \affiliation{School of Mathematical
Sciences, University of Electronic Science and Technology of China,
Chengdu 611731, China}

\author{Tatsuya Sasaki}
\affiliation{Faculty of Mathematics, University of Vienna, 1090
Vienna, Austria}

\author{Matja{\v z} Perc}
\affiliation{Faculty of Natural Sciences and Mathematics, University
of Maribor, SI-2000 Maribor, Slovenia}

\affiliation{Department of Physics, Faculty of Sciences, King
Abdulaziz University, Jeddah, Saudi Arabia}

\affiliation{Center for Applied Mathematics and Theoretical Physics,
University of Maribor, SI-2000 Maribor, Slovenia}

\begin{abstract}
Monitoring with implicated punishment is common in human societies to avert freeriding on common goods. But is it effective in promoting public cooperation? We show that the introduction of
monitoring and implicated punishment is indeed effective, as it transforms the public goods game to a coordination game, thus rendering cooperation viable in infinite and finite well-mixed populations. We also show that the addition of within-group enforcement further promotes the evolution of public cooperation. However, although the group size in this context has nonlinear
effects on collective action, an intermediate group size is least conductive to cooperative behaviour. This contradicts recent field observations, where an intermediate group size was declared optimal with the conjecture that group-size effects and within-group enforcement are responsible. Our theoretical research thus clarifies key aspects of monitoring with implicated punishment in human societies, and additionally, it reveals fundamental group-size effects that facilitate prosocial collective action.
\end{abstract}

\keywords{public cooperation, implicated punishment, monitoring, within-group enforcement, group-size effects}

\maketitle

Public cooperation is imperative for the sustainable
management of common resources in human societies
\cite{ostrom_90,nowak_06,poteete_10}. However, human cooperation is
threatened by temptations that are rooted in selfish but lucrative
short-term benefits on offer when defecting or free-riding on the
efforts of others \cite{rand_tcs13}. Like rewarding
\cite{dreber_n08,rand_s09,hilbe_prsb10,
szolnoki_epl10,szolnoki_prx13}, punishment is often employed for
maintaining sufficiently high levels of public cooperation
\cite{sigmund_pnas01,sigmund_tee07,raihani_tee12,yang_w_pnas13,przepiorka_prsb13,sasaki_prsb13,wang_ecology14,raihani_tee15,lee_jh_jtb15,wang_jae15}.
In addition to individual efforts aimed at punishing free-riders
\cite{fehr_n02,boyd_pnas03,fu_pre08,fu_pre09,shirado_nc13,fu_srep12},
our societies are home to a plethora of sanctioning institutions
\cite{gurerk_s06,sasaki_pnas12}. In particular, during the last
decade peer and pool punishment have been studied theoretically and
experimentally as possible means to stabilize cooperation
\cite{fehr_n02,gardner_a_an04,henrich_s06,
gurerk_s06,fowler_pnas05,helbing_ploscb10,boyd_pnas03,
egas_prsb08,hauert_s07,yamagishi_jpsp86,perc_srep12, sigmund_n10}.

Although ample research efforts have already been invested to inform
on the subtleties of positive and negative reciprocity and their
role in promoting public cooperation \cite{sigmund_tee07}, few
studies have thus far considered implicated punishment despite it
being and integral cog in various sanctioning systems in human
societies. In general, the implementation of implicated punishment
means that once a wrongdoer is caught, all the group members are
punished, no matter whether the group members are cooperators or
defectors. Such punishment schemes are particularly common for
monitoring \cite{rustagi_science10} the management of common
resources on large scales. For example, in Nature Reserve of China,
an administrative bureau is responsible for monitoring all illegal
activities. When the bureau staff members detect an illegal activity
in the monitored parcel, all households within the group will suffer
the same fine \cite{yang_w_pnas13}. While the system may work in
practice, in theory it is still unclear how fines affect cooperators
that are adversely affected, and how the overall dynamics plays out
in favor of prosocial behaviour.

In addition to the well-known and important adverse effects that
emerge if cooperators are sanctioned \cite{herrmann_s08, rand_nc11},
some individuals in the group may act emotionally and exploit
options related to within-group enforcement \cite{xiao_pnas05,
egas_prsb08, raihani_bl12, bone_bl14}, for example resorting to
probabilistic peer punishment \cite{chen_xj_njp14}. It is thus also
of interest to consider whether the addition of probabilistic
within-group enforcement can further enhance the evolution of
cooperation in the presence of monitoring and implicated punishment.
In fact, a recent study based on field observations found that an
intermediate group size is optimal for public cooperation when both
implicated punishment and within-group enforcement are present
\cite{yang_w_pnas13}. However, there is no theoretical research
available that would support the conjecture that group-size effects
and within-group enforcement are responsible for the success of
implicated punishment.

In this paper, we therefore consider a public goods game with
implicated punishment and within-group enforcement in infinite and
finite well-mixed populations. Our goal is to develop a thorough
theoretical understanding behind the success of implicated
punishment, and the role within-group enforcement and group size
play in either supporting or impairing the evolution of public
cooperation. As we will show, implicated punishment transforms the
public goods game into a coordination game, and within-group
enforcement further promotes the emergence of prosocial collective
action. Contrary to field observations \cite{yang_w_pnas13},
however, theory fails to predict an optimal intermediate group size
for the evolution of cooperation. Instead, we find that an
intermediate group size is actually not beneficial for the
successful evolution of cooperation. Our research thus clarifies key
aspects of monitoring with implicated punishment in human societies,
and it also reveals fundamental group-size effects that may promote
a public agenda.

\section*{Results}

\begin{figure}
\centering
\includegraphics[width=8.5cm]{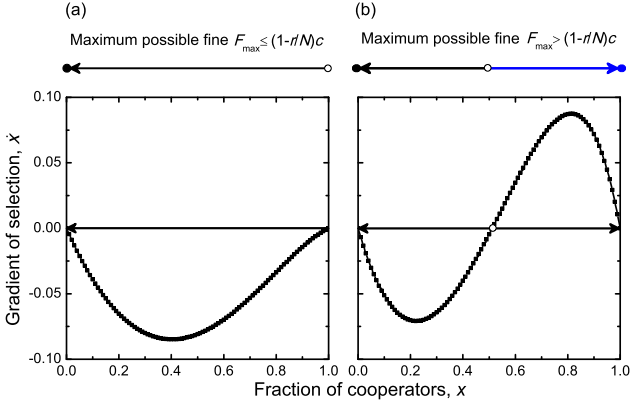}
\caption{The gradient of selection in dependence on the
fraction of cooperators in infinite populations. Stable equilibria
are depicted with solid circles, while unstable equilibria are
depicted with open circles. Arrows indicate the expected direction
of evolution. Cooperation is favored over defection when the arrow
points to the right. When the maximal possible average fine for a
defector $F_{max}=dp+pq(N-1)\beta\leq (1-r/N)c$, the public good
dilemma still exists with full defection as the only stable
equilibrium (a). Otherwise, the public good game is transformed into
a coordination game with full cooperation and full defection as the
two stable equilibria (b). Parameter values are: $N=5$, $r=3$,
$c=1$, $d=1.0$, $p=0.1$, $\alpha=0.3$, $\beta=1.0$, and $q=0.5$ in
(a); $N=5$, $r=3$, $c=1$, $d=1.0$, $p=0.5$, $\alpha=0.3$,
$\beta=1.0$, and $q=0.5$ in (b).} \label{fig1}
\end{figure}

We consider a well-mixed population, in which individuals
engage in a public goods game \cite{hauert_jtb02}, where each
individual is able to cooperate or to defect, respectively. In each
group, $N$ players are chosen randomly to form a group for playing
the game. Cooperators contribute the cost $c$, while defectors
contribute nothing. The sum of all contributions in the group is
multiplied by the enhancement factor $r>1$, and then split evenly
among all group members. After choosing the strategy, the group's
behaviours will be monitored with a probability $p$ $(0<p<1)$. If it
is detected that there is at least one defector in the group, then
the implicated punishment mechanism will work, and accordingly each
individual will incur a fine $d$ $(d>0)$. Otherwise, there is no
monitoring, and there is no fine on any individual. But once the
implicated punishment is implemented in the group, it may trigger
the within-group enforcement. Accordingly, each cooperator (if
present) will use the peer punishment on defectors with a
probability $q$ $(0<q<1)$, and is designated as a punisher. Peer
punishers impose a fine $\beta$ on each defector at a cost $\alpha$
$(0<\alpha<\beta$).

Below, we study how the introduction of implicated punishment and
within-group enforcement influences the evolutionary dynamics of
cooperation both in infinite and finite well-mixed population, in
particular the effects of group size in the model, by theoretical
and numerical analysis. We emphasize that the social dilemma only
exists when $r<N$ in the public goods game
\cite{hauert_jtb02,szabo_pr07}, so in this study the interval of $r$
values is constrained as $1<r<N$.

We first present the gradient of selection $\dot{x}$ given by the
replicator equation (Methods for infinite populations) for studying
the evolution of cooperative behaviour in infinite populations, as
illustrated in Fig.~\ref{fig1}. Here, $x$ is the fraction of all the
cooperators in the infinite population. We show that there exist two
typical behaviours for the gradient of selection varying with the
fraction of cooperators, as presented in Fig.~\ref{fig1}(a) and
Fig.~\ref{fig1}(b) respectively. We define $F_{max}$ as the maximal
fine upon a defector who receives from the two punishment regimes,
$dp+pq(N-1)\beta$. We accordingly prove that if $F_{max}\leq
(1-r/N)c$ (Methods for infinite populations), the gradient of
selection is always negative (Fig.~\ref{fig1}(a)). Cooperators thus
die out regardless of the initial conditions. While if
$F_{max}>(1-r/N)c$, a new unstable equilibrium emerges in the $x\in
(0, 1)$ interval, which divides the system into two basins of
attraction (Fig.~\ref{fig1}(b)). Depending on the initial
conditions, thus the system will evolve either towards full
defection or towards full cooperation. Both $x=0$ and $x=1$ are
stable steady states, indicating that the public goods game is
transformed into a coordination game.

\begin{figure}
\centering
\includegraphics[width=8.5cm]{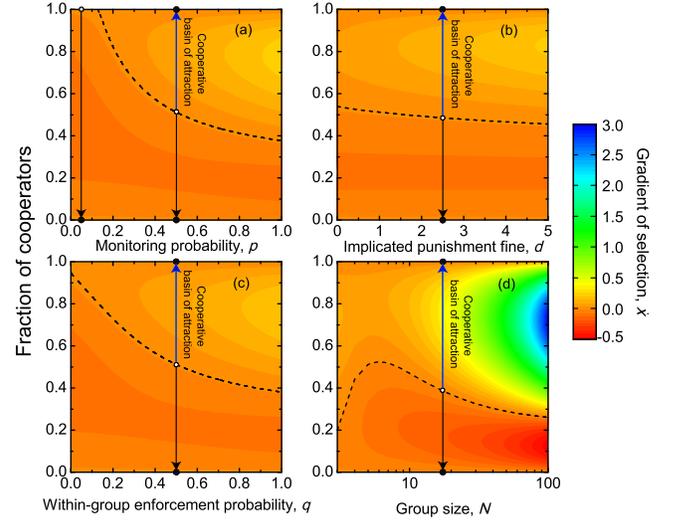}
\caption{The stationary fraction of cooperators and gradient
of selection in infinite populations. The unstable internal
equilibrium (if present) dividing the system into two basins of
attraction is indicated by dash line and also by open circles, and
the cooperative basin of attraction is indicated by blue arrows. The
stable boundary equilibrium is indicated by solid circles, while the
unstable boundary equilibrium is indicated by open circles. The
gradient of selection in the areas above the dash line is positive,
while the gradient of selection in the areas below the dash line is
negative. And the magnitude of the gradient of selection is shown
using the red-green-blue scale indicated, and blue areas indicate
parameter combination for which the fraction of cooperators increase
faster. In (a-c), the unstable internal equilibrium decreases with
increasing the monitoring probability $p$, the implicated punishment
fine $d$, and the within-group enforcement probability $q$,
respectively. While in (d) the unstable internal equilibrium first
increases, then decreases with increasing the group size $N$. In
other words, increasing $p$, $d$, or $q$ enlarges the basin of
attraction of the $x=1$ stable state, thus favoring the evolution of
cooperation. Importantly, a small group size or a large group size
can lead to a larger basin of attraction of the $x=1$ stable state
than an intermediate group size does. In addition, in the areas
above the dash lines, the rate of increase of the fraction of
cooperators depends on the parameter values. Parameter values are:
$N=5$, $r=3$, $c=1$, $d=1.0$, $\alpha=0.3$, $\beta=1.0$, and $q=0.5$
in (a); $N=5$, $r=3$, $c=1$, $p=0.5$, $\alpha=0.3$, $\beta=1.0$, and
$q=0.5$ in (b); $N=5$, $r=3$, $c=1$, $p=0.5$, $d=1.0$, $\alpha=0.3$,
and $\beta=1.0$ in (c); $r=3$, $c=1$, $p=0.5$, $d=1.0$,
$\alpha=0.3$, $\beta=1.0$, and $q=0.5$ in (d).} \label{fig2}
\end{figure}

\begin{figure}
\centering
\vspace{-0.31cm}
\includegraphics[width=8cm]{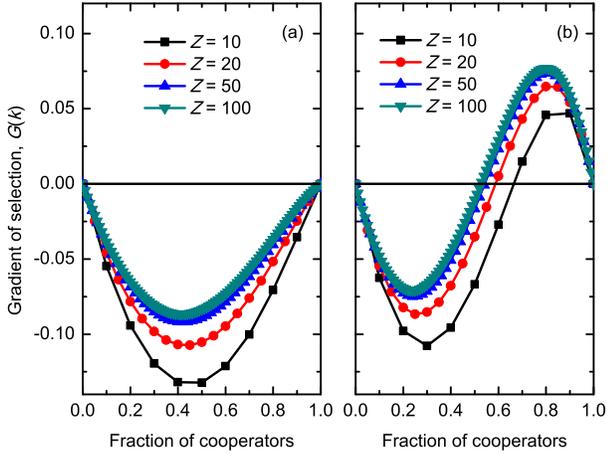}
\caption{The gradient of selection in dependence on the
fraction of cooperators in finite populations for different
population size. The gradient of selection exhibits two
qualitatively identical behaviours as reported in Fig.~\ref{fig1}.
Parameter values are: $N=5$, $r=3$, $c=1$, $p=0.1$, $d=1.0$,
$\alpha=0.3$, $\beta=1.0$, and $q=0.5$ in (a); $N=5$, $r=3$, $c=1$,
$p=0.5$, $d=1.0$, $\alpha=0.3$, $\beta=1.0$, and $q=0.5$ in (b).}
\label{fig3}
\end{figure}

Furthermore, we investigate how the parameters influence the
stationary fraction of cooperators in the infinite population, as
shown in Fig.~\ref{fig2}. We find that when the monitoring
probability $p$ is zero or small, there is always no interior
equilibrium, regardless of the values of other parameters in
Fig.~\ref{fig2}(a). When $p$ increases to $c(1-r/N)/[d+q(N-1)\beta]$
(Methods for infinite populations), an interior equilibrium which is
unstable enters the state space at the point $x=1$. With further
increasing $p$, the interior equilibrium decreases. In other words,
increasing the monitoring probability enlarges the basin of
attraction of the $x=1$ steady state. We now consider the effects of
implicated fine $d$. When $d=0$, if $pq(N-1)\beta>c(1-r/N)$, then
there is an interior equilibrium (Methods for infinite populations).
Otherwise, no interior equilibrium can emerge. If the interior
equilibrium is present, it decreases with increasing $d$
(Fig.~\ref{fig2}(b)), which means that increasing the implicated
fine $d$ also enlarges the basin of attraction of the $x=1$ steady
state. It is necessary to point out that compared to the increase of
$p$, the increase of $d$ makes the value of the interior equilibrium
decrease much slowly. This means that the chance of monitoring can
result in more positive effects on the evolution of cooperation than
the punishment fine does, when the probabilistic implicated
punishment is considered. In addition, when the probability for
within-group enforcement $q$ is zero, the interior equilibrium
presents if $dp>c(1-r/N)$ (Methods for infinite populations). Then
it decreases with increasing $q$, accordingly the basin of
attraction of the $x=1$ steady state is enlarged
(Fig.~\ref{fig2}(c)). Finally, we investigate the effects of group
size $N$. Interestingly, we find that if the interior equilibrium is
present, it first increases, reaches a maximum, but then decreases
with increasing the group size (Fig.~\ref{fig2}(d)). This means that
the basin of attraction of the $x=1$ steady state is smallest at an
intermediate group size. We also find that the interior equilibrium
could be absent for small group size, depending on the values of
other parameters. But it will exhibit then when the group size
increases to a certain value (Supplementary Fig.~S1). Subsequently,
the interior equilibrium decreases with further increasing the group
size, which indicates that the larger the group size, the greater
the basin of attraction of $x=1$. This finding is in agreement with
previous experimental results in \cite{isaac_jpe94}. Furthermore, we
emphasize that no matter how large the values of $p$, $d$, $q$, and
$N$ are, the boundary equilibrium $x=0$ is always stable, which
means that the outcome that $x=1$ is the only stable state cannot
happen in our model (Methods for infinite populations).

It is worth pointing out that in line with \cite{yang_w_pnas13},
group size is found to be able to produce nonlinear effects on
collective action in our study. But being contrary to the field
observation, we find that an intermediate group size cannot lead to
the most favorable outcome for public cooperation. Instead, it could
lead to the smallest basin of attraction of the full cooperation
state, which indicates that an intermediate group size is not
beneficial to the evolution of cooperation when the implicated
punishment and within-group enforcement are incorporated.

\begin{figure}
\centering
\includegraphics[width=8.5cm]{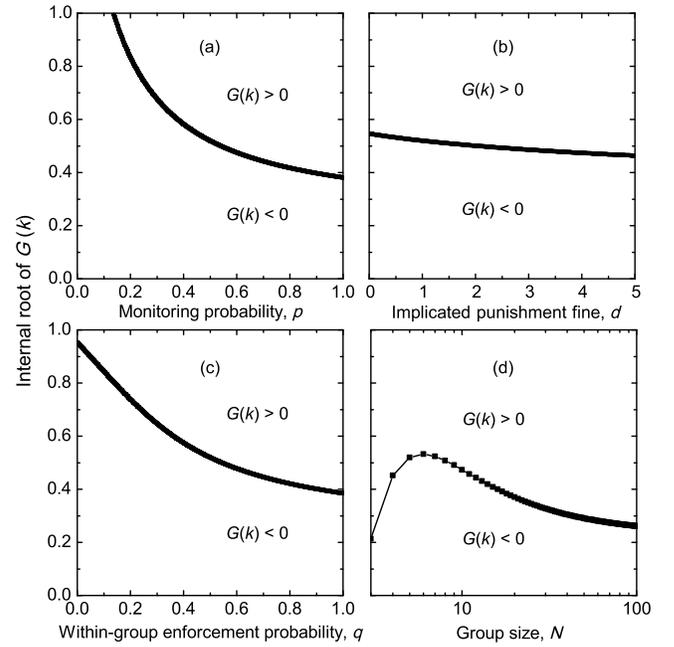}
\caption{The internal roots of the gradient of selection
$G(k)$ in finite populations. The roots are normalized by the
population size $Z$. In (a-c), the values of roots decrease with
increasing the monitoring probability $p$, the fine of implicated
punishment $d$, and the probability of within-group enforcement $q$,
respectively. While in (d), the values first increase, and then
decrease with increasing the group size $N$. Parameter values are:
$Z=200$, $r=3$, $c=1$, $N=5$, $d=1.0$, $\alpha=0.3$, $\beta=1.0$,
and $q=0.5$ in (a); $Z=200$, $r=3$, $c=1$, $N=5$, $p=0.5$,
$\alpha=0.3$, $\beta=1.0$, and $q=0.5$ in (b); $Z=200$, $r=3$,
$c=1$, $N=5$, $p=0.5$, $\alpha=0.3$, $\beta=1.0$, and $d=1.0$ in
(c); $Z=200$, $r=3$, $c=1$, $p=0.5$, $d=1.0$, $\alpha=0.3$,
$\beta=1.0$, and $q=0.5$ in (d).} \label{fig4}
\end{figure}

\begin{figure*}
\centering
\includegraphics[width=14.5cm]{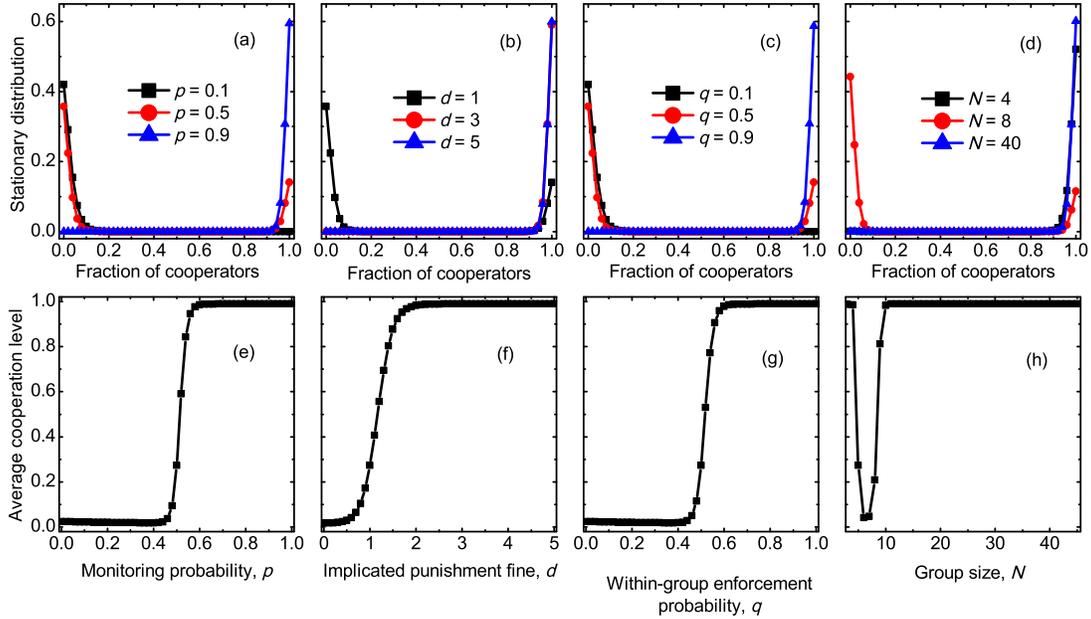}
\caption{The stationary distribution and the average
cooperation level. Top row depicts the stationary distribution in
finite populations in the presence of mutation $u=0.01$. In (a-c),
the population spends more time in states where cooperators prevail
for a larger monitoring probability $p$, a larger fine of implicated
punishment $d$, or a larger probability of within-group enforcement
$q$. While in (d) the population spends more time in states where
cooperators thrive for either a small group size or a large group
size, and spends less time in states where cooperators decline for
an intermediate group size. Bottom row depicts the average value of
cooperation level in the presence of mutation $u=0.01$. In (e-g),
the average cooperation level increases with increasing the
monitoring probability $p$, the fine of implicated punishment $d$,
or the probability of within-group enforcement $q$. While in (h),
the average cooperation level reaches a high value at either a small
group size or a large group size, and reaches a minimum at an
intermediate group size. Parameter values are: $Z=50$, $r=3$, $c=1$,
$N=5$, $d=1.0$, $\alpha=0.3$, $\beta=1.0$, and $q=0.5$ in (a) and
(e); $Z=50$, $r=3$, $c=1$, $N=5$, $p=0.5$, $\alpha=0.3$,
$\beta=1.0$, and $q=0.5$ in (b) and (f); $Z=50$, $r=3$, $c=1$,
$N=5$, $p=0.5$, $\alpha=0.3$, $\beta=1.0$, and $d=1.0$ in (c) and
(g); $Z=50$, $r=3$, $c=1$, $p=0.5$, $d=1.0$, $\alpha=0.3$,
$\beta=1.0$, and $q=0.5$ in (d) and (h).} \label{fig5}
\end{figure*}

In addition, we show the gradient of selection $\dot{x}$ in
Fig.~\ref{fig2}, and indicate that its value in the areas above the
dash line is positive. If the gradient of selection is positive, the
fraction of cooperators will increase. We see that with increasing
the monitoring probability $p$, the implicated punishment fine $d$,
or the within-group enforcement probability $q$, the gradient of
selection increases in the ares where $\dot{x}>0$. However, in that
area the gradient of selection first decreases, reaches a minimum,
but then increases with increasing the group size. But for a fixed
value of $p$, $d$, $q$, or $N$, the gradient of selection can always
reach the maximum values at an intermediate fraction of cooperators,
which is relatively smaller than $x=1$.

Corresponding to the right-hand side of the replicator equation, we
use the gradient of selection $G(k)$
\cite{pacheco_prsb09,souza_mo_jtb09} (Methods for finite
populations), to describe the behavioural dynamics in finite
populations. Figure~\ref{fig3} shows two typical behaviours of
$G(k)$ as a function of the fraction of cooperators $k/Z$ for
different sizes $Z$ of finite populations. We find that the two
typical behaviours found in infinite populations are also valid in
finite populations, for any parameter combinations. One behaviour
depicts that $G(k)<0$ for any $k$, which shows that cooperators are
always disadvantageous. The other depicts that $G(k)$ has a unique
internal root $k^*$, above which $G(k)>0$. This means that
cooperators become advantageous when $k$ is larger than $k^*$. In
addition, with increasing the population size, the gradient of
selection increases. This results in that the position of the
interior root moves from right to left by increasing the population
size. Thus, the range of $k/Z$ in which cooperators are advantageous
is greatly increased for large populations.

In what follows, we show how the interior root of $G(k)$ varies with
the parameters which have been referred to infinite populations
(Fig.~\ref{fig4}). We find that when the root exists, its value
monotonically decreases with increasing the monitoring probability
$p$, the implicated sanction fine $d$, or the within-group
enforcement probability $q$ (Supplementary Fig. S2(a-c)). This means
that the range of $k/Z$ for which cooperation is advantageous
increases when any one of these three parameters ($p$, $d$, and $q$)
increases. It is worth pointing out that the value of the interior
root decreases much slowly as the implicated punishment fine $d$
increases, and this phenomenon is also found in infinite populations
indicating that the punishment fine can only provide limited
positive effects on cooperation. While with increasing the group
size, the root's value first increases, reaches a maximum, but then
decreases again (Fig.~\ref{fig4}(d)). This means that the range of
$k/Z$ for which cooperation is advantageous reaches the minimal
value at an intermediate group size. However, the root may be
present only when the group size exceeds a certain value for other
parameter values (Supplementary Fig. S2(d)). With further increasing
the group size, the root monotonically decreases. Accordingly, the
obtained results in finite population confirm that an intermediate
group size is not optimal for the evolution of cooperation, but it
is certainly not detrimental for cooperation. This in turn indicates
that the combined effects of free-riding and within-group
enforcement do not lead to an optimal intermediate group size,
contrary to the conjecture in \cite{yang_w_pnas13}. In addition, we
emphasize that the root's value recovers to that in Fig.~\ref{fig2}
when $Z\rightarrow +\infty$, and the dependence of the root's value
on these parameters ($p$, $d$, $q$, and $N$) is very similar to
those in infinite well-mixed populations.

Another key quantity for describing the evolutionary dynamics in
finite well-mixed populations is the stationary distribution in the
presence of mutations (Methods for finite populations)
\cite{santos_pnas11,moreira_srep13}. In the top row of
Fig.~\ref{fig5}, we show how the stationary distribution changes
with the four parameters ($p$, $d$, $q$, and $N$), respectively. It
is worth pointing out that the stationary distribution characterizes
the pervasiveness in time of a given configuration of the
population. We find that with increasing the monitoring probability
$p$, the implicated sanction fine $d$, or the within-group
enforcement probability $q$, the time that the system spends in the
full cooperation state increases. With the large values of these
parameters, the system spends most of the time in the full
cooperation state, leading to maxima of the stationary distribution
at $k=Z$. But the time that the system spends in the full
cooperation state does not monotonically increase with increasing
the group size. Instead, with an intermediate group size, the system
spends most of the time in the full defection state, leading to
maxima of the stationary distribution at $k=0$. While either a small
group size or a large group size leads to that the system spends
most of the time in the full cooperation state.

In the bottom row of Fig.~\ref{fig5}, we further show how the
average value of cooperation level varies with the four parameters
($p$, $d$, $q$, and $N$), respectively. We find that the average
cooperation level monotonically increases with increasing the
monitoring probability $p$, the implicated sanction fine $d$, or the
within-group enforcement probability $q$. But we observe that with
increasing the group size, it first decreases, reach a minimum, then
increases again. This means that an intermediate group size is not
beneficial to the evolution of cooperation. Altogether,
Fig.~\ref{fig5} confirms that cooperation is promoted either at a
small group size or a large group size, rather than an intermediate
group size.

In the Supplementary Information, we also investigate our model in
finite populations with large peer punishment cost $\alpha$
(Supplementary Fig. S3), and explore the effects of the selection
intensity (Supplementary Fig. S4) and the mutation rates
(Supplementary Fig. S5) on the stationary distribution of
cooperation and the average cooperation level. We find that our
results regarding the effects of the monitoring probability, the
implicated punishment fine, the within-group enforcement
probability, and the group size are not changed when the above
variations are considered. In addition, we consider a discounting
factor for the implicated punishment fine on cooperators
(Supplementary Fig. S6). We find that the introduction of the
discounting factor does not change the genetic outcome about the
internal root of the gradient of selection in infinite and finite
populations, but can decrease the value of the internal root, thus
increasing the advantage of cooperators.

\section*{Discussion}
Human cooperation is unique, and it is one of the key
pillars of our evolutionary success. The origins of our remarkable
other-regarding abilities are likely rooted in the mitigation of
between-group conflicts, and in the necessity for alloparental care
during the advent of the genus \textit{Homo}. In the absence of such
pressing challenges, however, human societies rely on rewarding and
policing to maintain public cooperation \cite{ostrom_90}. Monitoring
with implicated punishment is a special form of policing, and this
form of monitoring and punishment is particularly common. In this
paper, based on an evolutionary game theoretical model we have
studied the monitoring with implicated punishment and within-group
enforcement in infinite and finite well-mixed populations.

As we have emphasized above, our model setup is well aligned with
reality in that implicated punishment and within-group enforcement
are common in human societies, and it is indeed relatively
straightforward to come up with examples where our model could
apply. A good example is the large-scale management of common
resources in general. The key assumption of implicated punishment is
that once a defector within a group is detected, subsequently all
members of that group, regardless of their strategies, are fined.
Evidently, it is thus likely that cooperators will be punished too.
As a countermeasure, we have also considered within-group
enforcement through peer punishment. We have shown that the
implicated punishment alone transforms the public goods game into a
coordination game. Accordingly, cooperation becomes viable, albeit
depending somewhat on initial conditions. Adding within-group
enforcement to the setup, we have shown that this further relaxes
the necessary conditions for coordinated action to emerge, and thus
for public cooperation to thrive. Moreover, we have confirmed that
cooperation can be enhanced both in infinite and finite well-mixed
populations, thus establishing for the first time mechanisms that
underlie the success of implicated punishment. Our results also
indicate that in the probabilistic implicated punishment the fine
has an effect earlier than the monitoring probability for the
evolution of cooperation, but before any monitoring benefits
materialize a sufficient non-zero punishment fine is required. We
hope that this indication about the effects of the punishment fine
and the monitoring probability could be helpful for the policy
recommendations in the management of common resources.

Since the group size has been identified as a crucial factor
affecting collective action
\cite{bonacich_jcr76,isaac_qje88,carpenter_geb07,pecorino_jpet08,szolnoki_pre11c},
we have also considered this aspect of the studied evolutionary game
in detail. In the typical public goods game, the negative effect of
free-riding on cooperation are enhanced by increasing the group
size. But when punishment is introduced into the game, it has a
positive effect on cooperation especially for large group size
\cite{carpenter_geb07}. The coexistence of these two opposing
factors determines the net effect of the group size, and ultimately
the combination of free-riding and punishment leads to the group
size having nonlinear effects on collective behaviour. This is in
fact predicted quantitatively by our theoretical analysis, and is in
agreement with a recent field investigation involving free-riding
and within-group enforcement \cite{yang_w_pnas13}. However, the
difference is that our theoretical results show that an intermediate
group size is not best for cooperative behaviour, while the field
data show the opposite. Importantly, while the conclusions of the
field investigation rely solely on the effects of free-riding and
within-group enforcement, and also because the range of the
available group sizes in the field data was small
\cite{yang_w_pnas13}, the two opposing factors predicted by our
theoretical analysis could not have been taken into account. Our
study thus provides further key insights on the intricate interplay
between the group size, within-group enforcement, and implicated
punishment. We hope that our in part counterintuitive results will
inspire further theoretical and empirical research devoted to the
mechanisms that are essential for prosocial collective behaviour.

\section*{Methods}

\subsection*{Evolutionary dynamics in infinite well-mixed populations}
For studying the evolutionary dynamics in infinite well-mixed
populations, we use the replicator equation \cite{hofbauer_98}. To
begin, we assume a large population, a fraction $x$ of which is
composed of cooperators, the remaining fraction $(1-x)$ being
defectors. Accordingly, the replicator equation is
\begin{equation}
\dot{x}=x(1-x)(P_X-P_D),\label{eq1}
\end{equation}
where $P_X=qP_P+(1-q)P_C$ is the average payoff of all the
cooperators, while $P_P$, $P_C$, and $P_D$ are the average payoffs
of punishing cooperators, second-order free riders (cooperators who
do not punish), and defectors, respectively. And the average payoffs
$P_C$, $P_P$, and $P_D$ are respectively
\begin{widetext}
\begin{eqnarray*}
P_C &=&\sum_{i=0}^{N-2}\left(\begin{array}{c}
N-1\\i\end{array}\right)x^i(1-x)^{N-1-i}[(1-p)(\frac{i+1}{N}rc-c)+p(\frac{i+1}{N}rc-c-d)]+x^{N-1}(rc-c)
\nonumber \\
&=&\frac{rc}{N}+\frac{(N-1)rcx}{N}-c-pd+dpx^{N-1},
\end{eqnarray*}

\begin{eqnarray*}
P_P
&=&\sum_{i=0}^{N-2}\left(\begin{array}{c}N-1\\i\end{array}\right)x^i(1-x)^{N-1-i}\{(1-p)(\frac{i+1}{N}rc-c)+p[\frac{i+1}{N}rc-c-d-(N-1-i)\alpha]\}
\nonumber
\\&&+x^{N-1}(rc-c)\nonumber \\ &=&\frac{rc}{N}+\frac{(N-1)rcx}{N}-c-dp+p\alpha (N-1)(x-1)+dpx^{N-1},\,\,  {\rm and}
\end{eqnarray*}

\begin{eqnarray*}
 P_D &=& \sum_{i=0}^{N-1}\left (
\begin{array}{c} N-1\\i\end{array} \right)x^i(1-x)^{N-1-i}\cdot[(1-p)\frac{i}{N}rc+p\sum_{j=0}^{i}\left(\begin{array}{c}
i\\j\end{array}\right)q^j(1-q)^{i-j}(\frac{i}{N}rc-d-j\beta)] \nonumber \\
 &=& \frac{(N-1)rcx}{N}-dp-pq\beta(N-1)x,
\end{eqnarray*}
\end{widetext}
where $i$ denotes the number of all the cooperators among $N-1$
co-players in a group, and $j$ ($j\leq i$) denotes the number of
punishing cooperators among $i$ cooperators.

With these definitions, the replicator equation has two boundary
equilibria, namely $x=0$ and $x=1$. Interior equilibria, on the
other hand, can be determined by the roots of the function
$g(x)=P_X-P_D$, thus obtaining
\begin{equation}
g(x)=dpx^{N-1}+pq(N-1)(\alpha+\beta)x-pq(N-1)\alpha-c+\frac{rc}{N}.
\end{equation}
It follows that $g(0)=-c+\frac{rc}{N}-pq(N-1)\alpha<0$ when $r<N$.
On the other hand, the function $g(x)$ is strictly increasing since
$g'(x)=dp(N-1)x^{N-2}+pq(N-1)(\alpha+\beta)>0$ for $x\in (0, 1)$.
Accordingly, the interior equilibrium is determined by
$g(1)=dp+pq(N-1)\beta-c+\frac{rc}{N}$, from which we have the
following two conclusions:

($1$) When $dp+pq(N-1)\beta>c(1-\frac{r}{N})$, the replicator
equation has only one interior equilibrium $x^{*}\in (0, 1)$, but it
is unstable since $g'(x^{*})>0$. The two boundary equilibria $x=0$
and $x=1$ are both stable.

($2$) When $dp+pq(N-1)\beta\leq c(1-\frac{r}{N})$, the replicator
equation has no interior equilibria in $(0, 1)$. Only $x=0$ is a
stable equilibrium, while $x=1$ is an unstable equilibrium.

\subsection*{Evolutionary dynamics in finite well-mixed populations}
For studying the evolutionary dynamics in finite well-mixed
 populations, we consider a population of finite size $Z$. Here, the
average payoffs of second-order free-riders, punishing cooperators,
and defectors in the population with $k$ cooperators are
respectively given by
\begin{widetext}
\begin{eqnarray*}
f_C(k)&=&\left (
\begin{array}{c} Z-1\\N-1\end{array} \right)^{-1}\sum_{i=0}^{N-2} \left (
\begin{array}{c} k-1\\i\end{array} \right) \left (
\begin{array}{c} Z-k\\N-i-1\end{array} \right)[(1-p)(\frac{i+1}{N}rc-c)+p(\frac{i+1}{N}rc-c-d)]\nonumber
\\&&+\left (
\begin{array}{c} Z-1\\N-1\end{array} \right)^{-1}\left (
\begin{array}{c} k-1\\N-1\end{array} \right)(rc-c) \nonumber
\\&=& \frac{rc}{N}[1+(k-1)\frac{N-1}{Z-1}]-c-dp+dp\left (
\begin{array}{c} Z-1\\N-1\end{array} \right)^{-1}\left (
\begin{array}{c} k-1\\N-1\end{array} \right),
\end{eqnarray*}

\begin{eqnarray*}
f_P(k)&=&\left (
\begin{array}{c} Z-1\\N-1\end{array} \right)^{-1}\sum_{i=0}^{N-2} \left (
\begin{array}{c} k-1\\i\end{array} \right) \left (
\begin{array}{c} Z-k\\N-i-1\end{array} \right)\{(1-p)(\frac{i+1}{N}rc-c)\nonumber
\\&& +p[\frac{i+1}{N}rc-c-d-(N-1-i)\alpha]\}+\left (
\begin{array}{c} Z-1\\N-1\end{array} \right)^{-1}\left (
\begin{array}{c} k-1\\N-1\end{array} \right)(rc-c) \nonumber
\\&=&
\frac{rc}{N}[1+(k-1)\frac{N-1}{Z-1}]-c-\frac{N-1}{Z-1}(Z-k)p\alpha-dp+dp\left
(
\begin{array}{c} Z-1\\N-1\end{array} \right)^{-1}\left (
\begin{array}{c} k-1\\N-1\end{array} \right),\,\,  {\rm and}
\end{eqnarray*}

\begin{eqnarray*}
f_D(k)&=&\left (
\begin{array}{c} Z-1\\N-1\end{array} \right)^{-1}\sum_{i=0}^{N-1} \left (
\begin{array}{c} k\\i\end{array} \right) \left (
\begin{array}{c} Z-1-k\\N-1-i\end{array} \right)\sum_{j=0}^{i} \left(\begin{array}{c}
i\\j\end{array}\right)q^j(1-q)^{i-j}[(1-p)\frac{i}{N}rc+p(\frac{i}{N}rc-d-j\beta)]\nonumber
\\&=&\frac{rc(N-1)}{N(Z-1)}k-dp-\frac{pq\beta (N-1)}{(Z-1)}k,
\end{eqnarray*}
\end{widetext}
where we impose that the binomial coefficients satisfy $\left (
\begin{array}{c} k-1\\N-1\end{array} \right)=0$ if $k<N$.

Consequently, the average payoff of all the cooperators is
\begin{eqnarray*} f_X(k)&=&qf_P(k)+(1-q)f_C(k)
\\&=&\frac{rc}{N}[1+(k-1)\frac{N-1}{Z-1}]-c-\frac{N-1}{Z-1}(Z-k)pq\alpha
\nonumber
\\&&-dp+dp\left (
\begin{array}{c} Z-1\\N-1\end{array} \right)^{-1}\left (
\begin{array}{c} k-1\\N-1\end{array} \right).
\end{eqnarray*}

Next, we adopt the pair-wise comparison rule to study the
evolutionary dynamics, based on which we assume that player $y$
adopts the strategy of player $z$ with a probability given by the
Fermi function
\begin{equation}
\frac{1}{1+\exp[-s(P_z-P_y)]},
\end{equation}
where $s$ is the intensity of selection that determines the level of
uncertainty in the strategy adoption process \cite{szabo_pre98,
szabo_pr07}. Without loosing generality, we use $s=2.0$ in
Fig.~\ref{fig5}, Supplementary Fig.~S3, and Supplementary Fig.~S5.

With these definitions, the probability that the number of
cooperators in the population increases or decreases by one is
\begin{equation}
T^{\pm}(k)=\frac{k}{Z}\frac{Z-k}{Z}[1+e^{\mp
s[f_X(k)-f_D(k)]}]^{-1}.
\end{equation}

In finite populations, the gradient of selection for arbitrary $s$
is thus given by
\begin{equation}
G(k)\equiv
T^{+}(k)-T^{-}(k)=\frac{k}{Z}\frac{Z-k}{Z}\tanh\{\frac{s}{2}[f_X(k)-f_D(k)]\}.
\end{equation}

We further introduce the mutation-selection process into the update
rule by assuming that mutations occur between cooperators and
defectors with probability $\mu$ in each update step
\cite{santos_pnas11,van-segbroeck_prl12}, and compute the stationary
distribution as a key quantity that determines the evolutionary
dynamics in finite well-mixed populations. We note that, in the
presence of mutations, the population will never fixate in any of
the two possible absorbing states. Thus, the transition matrix of
the complete Markov chain is
\begin{equation}
{\bf{M}}=[p_{m,n}]^{T},
\end{equation}
where $p_{m,n}=0$ if $|m-n|>1$,
$p_{m,m+1}=(1-\mu)T^{+}(m)+\mu(Z-m)/Z$,
$p_{m,m-1}=(1-\mu)T^{-}(m)+\mu m/Z$, and
$p_{m,m}=1-p_{m,m+1}-p_{m,m-1}$ otherwise. Accordingly, the
stationary distribution of the population, that is, the average
fraction of time the population spends in each of the $Z+1$ states,
is given by the eigenvector of the eigenvalue $1$ of the transition
matrix $\bf{M}$ \cite{karlin_75}. Specially, the unitized
eigenvector ${\bf{\Pi}}=[\pi_1,\cdot \cdot \cdot, \pi_{M+1}]^{T}$ is
derived explicitly for $l=1, \cdot \cdot \cdot, M+1$ \cite{ewns_04}:
$$\pi_l=\frac{\lambda_l}{\sum_{m=1}^{M+1} \lambda_m},$$
where $\lambda_l=1$ if $l=1$, and
$\lambda_l=\prod_{m=1}^{l-1}\frac{p_{m,m+1}}{p_{m+1,m}}$ otherwise.

In addition, another central quantity is the average cooperation
level $\bar{c}$, averaging over all possible states, weighted with
the corresponding stationary distribution
\cite{vasconcelos_ncc13,vasconcelos_pnas14}. Accordingly, $\bar{c}$
is computed as
$$\bar{c}={\bf{S}} {\bf{\Pi}}/Z,$$ where ${\bf{S}}=[0,\cdot \cdot \cdot,
Z]$ is the vector of population states.

\begin{acknowledgments}
This research was supported by the National Natural Science
Foundation of China (Grant No. 61503062), by the Fundamental
Research Funds of the Central Universities of China, by the Austrian
Science Fund (Grant P27018-G11), by the Deanship of Scientific
Research, King Abdulaziz University (Grant 76-130-35-HiCi), and by
the Slovenian Research Agency (Grant P5-0027).
\end{acknowledgments}

\end{document}